\documentclass[a4paper,11pt]{article}
\usepackage{jheppub}
\usepackage{lineno}
\usepackage{algorithm}
\usepackage{multirow}
\usepackage{algpseudocode}
\usepackage{amsmath}
\usepackage{amssymb}
\usepackage{graphicx}
\usepackage[normalem]{ulem}
\usepackage{soul}
\newcommand{\el}{\vspace*{0.2cm}}

\title{Physics-Conditioned Diffusion Models for Lattice Gauge Theory}

\author[a,b]{Qianteng Zhu}
\affiliation[a]{State Key Laboratory of Dark Matter Physics, Key Laboratory for Particle Astrophysics and Cosmology (MOE),
Shanghai Key Laboratory for Particle Physics and Cosmology,
Shanghai Jiao Tong University, Shanghai 200240, China}
\affiliation[b]{RIKEN Interdisciplinary Theoretical and Mathematical Sciences (iTHEMS), Wako, Saitama 351-0198, Japan}
\emailAdd{zhuqianteng@sjtu.edu.cn}

\author[c]{, Gert Aarts}
\emailAdd{g.aarts@swansea.ac.uk}
\affiliation[c]{Department of Physics, Swansea University, SA2 8PP, Swansea, United Kingdom}

\author[a,d]{, Wei Wang}
\affiliation[d]{Southern Center for Nuclear-Science Theory (SCNT), Institute of Modern Physics, Chinese Academy of Sciences, 96 South Sihuan Rd. Huicheng District, Huizhou 516000, Guangdong, China}
\emailAdd{wei.wang@sjtu.edu.cn}

\author[e,f]{, Kai Zhou}
\affiliation[e]{School of Science and Engineering, The Chinese University of Hong Kong (Shenzhen), Longgang, Shenzhen, Guangdong, 518172, P.R. China}
\affiliation[f]{School of Artificial Intelligence, The Chinese University of Hong Kong (Shenzhen), Longgang, Shenzhen, Guangdong, 518172, P.R. China}
\emailAdd{zhoukai@cuhk.edu.cn}

\author[b,g]{and Lingxiao Wang}
 \affiliation[g]{Institute for Physics of Intelligence, The University of Tokyo, Hongo,  Tokyo 113-0033, Japan}
\emailAdd{lingxiao.wang@riken.jp}

\abstract{We develop diffusion models for simulating lattice gauge theories, where stochastic quantization is explicitly incorporated as a physical condition for sampling. We demonstrate the applicability of this novel sampler to U(1) gauge theory in two spacetime dimensions and find that a model trained at a small inverse coupling constant can be extrapolated to larger inverse coupling regions without encountering the topological freezing problem. Additionally, the trained model can be employed to sample configurations on different lattice sizes without requiring further training. The exactness of the generated samples is ensured by incorporating Metropolis-adjusted Langevin dynamics into the generation process. Furthermore, we demonstrate that this approach enables more efficient sampling of topological quantities compared to traditional algorithms such as Hybrid Monte Carlo and Langevin simulations.}

\begin{document}
\maketitle

\flushbottom
\section{Introduction}

Lattice field theory is a discretization method for quantum field theory in which fields are defined on a spacetime lattice. The fundamental idea is to employ the Euclidean path integral formulation, representing the behaviour of quantum fields as a weighted sum over all possible field configurations. Observables are then defined as
\begin{equation}
\langle O \rangle = \frac{1}{Z} \int \mathcal{D}\phi \, O[\phi] e^{-S_E[\phi]}, \qquad Z = \int \mathcal{D}\phi \, e^{-S_E[\phi]},
\end{equation}
where $Z$ is the partition function, $S_E[\phi]$ is the discretized Euclidean action of the field, and $\phi$ represents a field variable $\phi(x)$ defined at spacetime position $x$.

Direct computation of path integrals is often challenging, particularly in higher dimensional systems or strongly-coupled regimes. To achieve an unbiased estimation of $\langle O \rangle$, traditional numerical simulation methods, such as the Markov Chain Monte Carlo (MCMC) algorithm, sample field configurations according to the probability distribution,

\begin{equation}
p[\phi] = \frac{1}{Z} e^{-S_E[\phi]}.
\end{equation}
Despite the success of MCMC in simulating lattice field theories, one often encounters the issue of critical slowing down \cite{Wolff:1989wq}. This phenomenon occurs when the correlation time of a system increases significantly as it approaches a critical point in parameter space, leading to escalating computational costs and insufficient diversity in the sampled configurations. Although alternative methods \cite{Wolff:1988uh,Hasenbusch:2001ne,Hasenbusch:2002ai,Prokofev:2001ddj,Brannick:2007ue,Albergo:2019eim,Albandea:2021lvl,Eichhorn:2023uge} have been proposed to mitigate critical slowing down in specific models, it remains a longstanding challenge in lattice simulations of Quantum Chromodynamics.

Generative models in machine learning (ML), which aim to map samples from a simple prior distribution to a complex physical distribution, have shown great promise in addressing challenges such as critical slowing down in lattice field theory. Recent advancements in generative models have introduced novel approaches for lattice field theory simulations~\cite{Cranmer:2023xbe,Zhou:2023pti,Wang:2020hji}. Among these, flow-based models have emerged as a leading method for explicit likelihood estimation, offering advantages such as invertibility and gauge equivariance, which enable efficient global sampling~\cite{Kanwar:2020xzo,Boyda:2020hsi,Kanwar:2021wzm,Albergo:2021vyo,Cranmer:2023xbe,Kanwar:2024ujc,Chen:2022ytr}. Furthermore, extensions of normalizing flows, including continuous normalizing flows~\cite{chen:2018,deHaan:2021erb,Gerdes:2022eve,Caselle:2023mvh} and stochastic normalizing flows~\cite{wu2020stochastic,Caselle:2022acb,Bulgarelli:2024brv}, expand their capabilities in lattice simulations. However, in flow-based models, the training complexity increases significantly with larger lattice sizes~\cite{DelDebbio:2021qwf,Abbott:2022zsh,Komijani:2023fzy,Abbott:2023thq}. Moreover, the use of the Kullback-Leibler divergence as the optimization objective can lead to mode collapse \cite{Nicoli:2023qsl,Kanaujia:2024zrq}, where the learned distribution fails to capture the full complexity of the target distribution. These challenges underscore the need to explore alternative generative models to further enhance sampling efficiency and reliability. 

Recently, diffusion models (DMs) have gained significant attention for their success in generating high-quality samples across diverse domains~\cite{sohl-dickstein:2015deep,ho:2020denoising,song2019generative,song2020score,yang2023diffusion,croitoru2023diffusion}, including applications in high-energy physics~\cite{Mikuni:2022xry,Amram:2023onf,Mikuni:2023dvk,Devlin:2023jzp}. Diffusion models iteratively denoise random noise into realistic configurations through probabilistic steps~\cite{bishop2023deep}. Their application to lattice field theory was pioneered in Refs.~\cite{Wang:2023exq,Wang:2023sry} for scalar fields, establishing the connection with stochastic quantization \cite{Parisi:1980ys,Damgaard:1987rr,Namiki:1993fd,Fukushima:2024oij}. Subsequent studies introduced a Feynman path integral formulation~\cite{Hirono:2024zyg} and analysed higher-order cumulants ~\cite{Aarts:2024rsl}, further solidifying the theoretical foundation of diffusion models in this context. An application to data generated using complex Langevin dynamics can be found in Ref.~\cite{Habibi:2024fbn}. In this work, we propose a novel framework based on diffusion models to address challenges in lattice gauge theory simulations~\cite{Zhu:2024kiu}. To validate our approach, we conduct experiments in two-dimensional U(1) gauge theory, demonstrating that our model effectively samples topological quantities across varying coupling constants and lattice sizes without requiring additional training. 

The paper is organized as follows. In 
Section~\ref{sec:lattice_gauge_fields} we discuss the issue of topological freezing, with subsections covering the Hybrid Monte Carlo (HMC) approach and topological charge. Section~\ref{sec:diffusion_models} introduces diffusion models and their training objectives, including the noise conditional score network and the Net architecture. In Section~\ref{sec:physics_conditioned_sampler}, we present our physics-conditioned sampler, which integrates the diffusion model with the notion of stochastic quantization and Langevin dynamics for conditional sampling while ensuring exactness. Section~\ref{sec:numerical_experiments} details our numerical experiments, including data preparation, model setups, and evaluations across various lattice sizes and coupling constants.  Section~\ref{sec:outlook} concludes. More details are discussed in the Appendices.

\section{Simulation of Lattice Gauge Fields}
\label{sec:lattice_gauge_fields}

Abelian gauge theories in two Euclidean dimensions represent quantum electrodynamics (QED) on a plane and provide simplified models for gauge field theories~\cite{Gattringer:2010zz,Greensite:2011zz,Kanwar:2020xzo,Crean:2024nro}, while retaining some of the essential features of nonabelian theories in four dimensions: fermion confinement~\cite{Coleman:1975pw} and separated topological sectors~\cite{Greensite:2011zz}. In the two-dimensional U(1) gauge theory, a gauge-invariant lattice Euclidean action can be defined as
\begin{equation}
    S_E = -\beta \sum_{x} \text{Re}(U_{x,\Box}),\label{eq:action}
\end{equation}
where the plaquette variable is
\begin{equation}
    U_{x,\Box} \equiv U_{x,\mu} U_{x+\hat{\mu},\nu} U_{x+\hat{\nu},\mu}^\dagger U_{x,\nu}^\dagger.
\end{equation}
On an $L \times L$ square lattice with periodic boundary conditions, a gauge field on lattice site $x$ in the direction $\mu$ is represented by a link variable $U_{x,\mu} = e^{i \phi_{x,\mu}}$, where $\phi_{x,\mu}$ is the gauge angle for the link. The plaquette, or the $1\times1$ Wilson loop ($W_{1\times1}$), is a gauge-invariant observable.
Here $\beta$ is the inverse coupling that controls the gauge interaction strength and ensures consistency with the continuum limit (at large $\beta$) in simulations~\cite{Gattringer:2010zz}.

\subsection{Hybrid Monte Carlo}
In lattice field theory, physical properties are estimated via the expectation value of observables $\hat{O}$, representing the statistical expectation over all configurations in the physics distribution. The HMC algorithm is widely used to sample configurations efficiently by introducing auxiliary momenta, and utilizing Hamiltonian dynamics~\cite{Duane:1987de,Hasenbusch:2001ne,Hasenbusch:2002ai,Gattringer:2010zz,Neal:2011mrf}. This enables effective phase space exploration and reduces autocorrelations, overcoming inefficiencies of random walk-based methods.

In particular, HMC evolves the system using Hamiltonian dynamics combined with a Metropolis acceptance step. The Hamiltonian $H$ consists of the gauge action $S[\phi]$ and the kinetic term for the conjugate momenta $\pi_{\mu}(x)$ as,
\begin{equation}
\label{eq:H}
H =  \frac{1}{2} \sum_{\mu, x} \pi^2_{\mu}(x) + S_E[\phi].
\end{equation}
One initializes the gauge fields $U_{\mu}(x)$, using the angles $\phi_{x,\mu}$ as degrees of freedom, randomly. For each gauge link, the conjugate momentum $\pi_{\mu}(x)$ is drawn from a Gaussian distribution. The gauge links and momenta are then evolved according to classical equation of motion obtained from Eq.~(\ref{eq:H}). These equations are discretized using a leapfrog integrator and, after a predefined number of molecular dynamics steps, we calculate the new Hamiltonian $H'$ and apply the acceptance criterion, $P_{\text{accept}} = \min(1, \exp{\{H - H'\}}).$ After generating a series of configurations, observables like  Wilson loops or the topological charge are estimated by averaging over the ensemble of configurations.

\subsection{Topological Charge Freezing}

In gauge theories, the topological charge plays a key role in phenomena like confinement~\cite{Greensite:2011zz}. It is quantized, characterizing different vacuum states, and is linked to flux quantization and instantons~\cite{Greensite:2011zz}. The Atiyah-Singer Index Theorem~\cite{Atiyah:1968mp} states that $Q$ must be an integer and remain unchanged under continuous field transformations. Thus, topological charges reside in distinct sectors separated by energy barriers~\cite{Eichhorn:2023uge,Rouenhoff:2024dal}.

In the two-dimensional case, the topological charge $Q$ indicates the winding number of the gauge field configuration and on the lattice is given by
\begin{equation}
    Q = \sum_x q(x) = \frac{1}{2\pi} \sum_x \arg U_{12}(x),
\end{equation}
where the phase of the plaquette is chosen within the principal interval, i.e., $\in (-\pi, \pi]$. The fluctuation of topological charge is described by the topological susceptibility, $\chi_Q = \langle Q^2\rangle/V$, where $V = L^2$ is the two-dimensional volume. 

\begin{figure}[hbpt!]
    \begin{center}
    \includegraphics[width=0.8\textwidth]{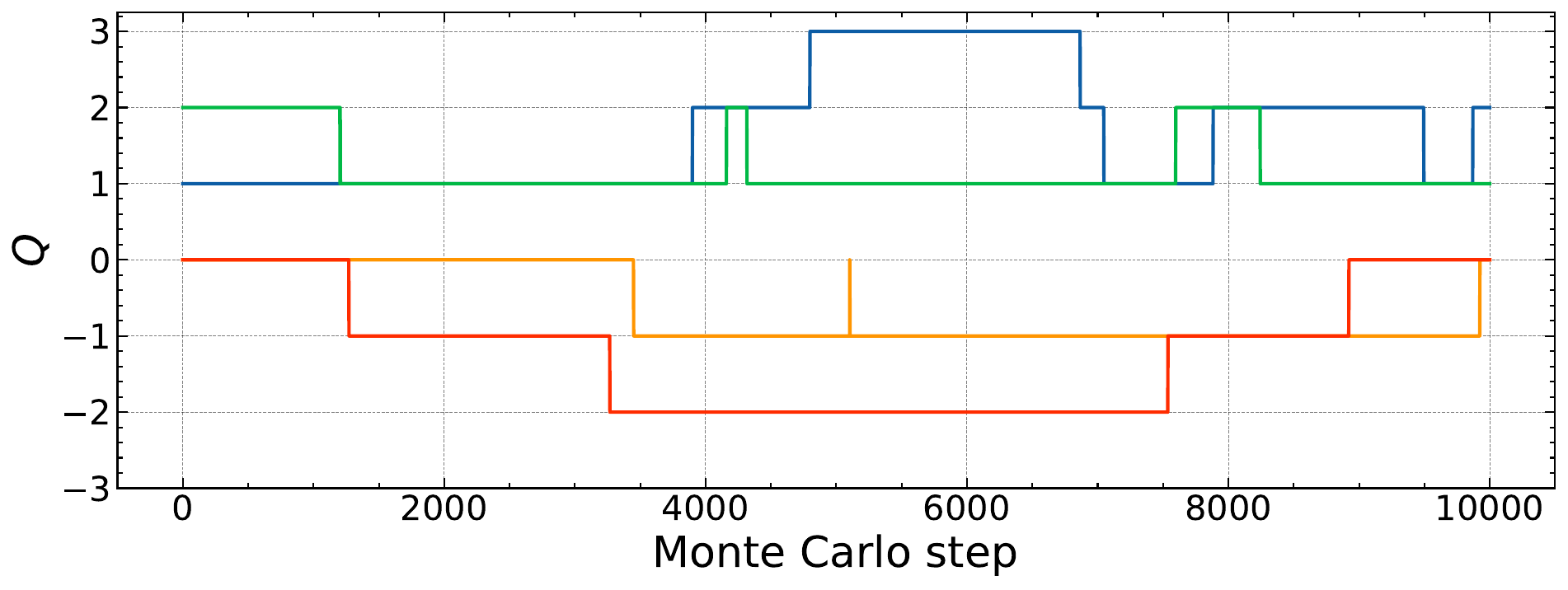}
    \caption{Monte Carlo evolution of the topological charge $Q$ on multi-Markov chains after thermalization with $L=16$ and $\beta=7$.}
    \label{fig:topof}
    \end{center}
\end{figure}

\textit{Topological freezing} refers to the phenomenon where the topological charge $ Q $ remains constant over long Monte Carlo trajectories, preventing adequate sampling of different topological sectors~\cite{Kanwar:2020xzo,Albandea:2021kwe}. This issue may arise as a general feature in Markov Chain Monte Carlo simulations of lattice theories with non-trivial topological properties~\cite{Alles:1996vn,DelDebbio:2002xa,DelDebbio:2004xh,Bonanno:2018xtd,Eichhorn:2023uge}. It is refereed to divergence of the autocorrelation time $\tau_Q$. Consequently, transitions between topological sectors become exponentially rare.

As illustrated in Fig.~\ref{fig:topof}, see also Figs.~1 of Refs.~\cite{Kanwar:2020xzo,Albandea:2021kwe}, samples within the same Markov chain (with different chains represented by different colors) remain confined to a limited number of topological sectors. This significantly hinders simulation efficiency and the accuracy of physical observables related to topology. In traditional Markov chain Monte-Carlo methods, when the system is trapped in a restricted region of configuration space, transitions between isolated topological sectors are further suppressed~\cite{Alles:1996vn,DelDebbio:2001sj}, situations exist even in hybrid methods like HMC, which is the most common way to sample QCD configurations. Specifically at large $\beta$, the action landscape may develop deep, narrow wells corresponding to these separated sectors, but Markov chains, which typically evolve by making small updates to state variables, struggle to transition between them~\cite{2001Annealed,Eichhorn:2023uge}. To address this challenge, improvements to existing algorithms or alternative approaches are necessary.

\section{Diffusion Model and Score Matching}
\label{sec:diffusion_models}
\subsection{Diffusion Models}
Diffusion models generate samples by reversing the process of adding noise, with the goal of recovering the original distribution from pure noise~\cite{bishop2023deep}. While theoretically possible if the forward stochastic dynamics are known, in practice it is often difficult to determine the reverse process. To solve this problem, a deep model approximates the gradient of the logarithm of the data distribution, which feeds directly in the denoising process. The well-trained deep model on prepared lattice field configurations can serve as an effective drift term to generate more field configurations by evolving a stochastic differential equation~\cite{Wang:2023exq}. It can provide an efficient and global sampler to simulate lattice quantum field theories~\cite{Wang:2023exq,Wang:2023sry,Aarts:2024rsl,Aarts:2024agm}.
Here, a \textit{global sampler} refers to a sampling method capable of proposing new configurations that differ globally from previous ones, effectively reducing the issue of long autocorrelation times and topological freezing commonly encountered in local update algorithms.

\begin{figure}[!htbp]
\begin{center}
\includegraphics[width=1.0\textwidth]{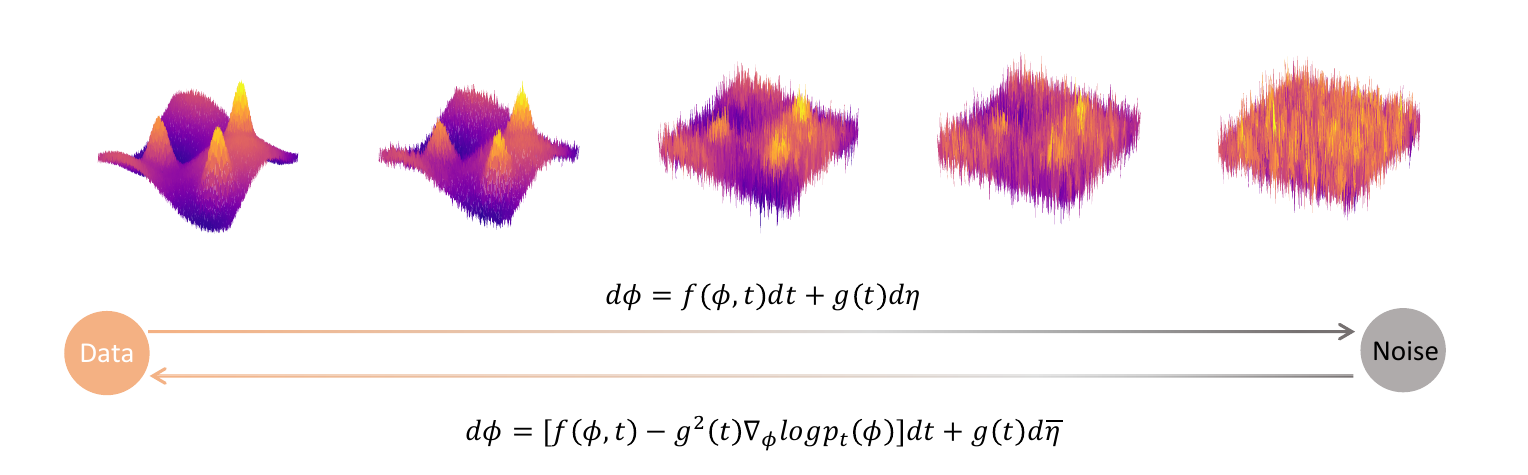}
\caption{The forward diffusion process gradually adds noise and the reverse denoising process tries to remove noise. The two stochastic processes are described by two stochastic differential equations. The target distribution is typically unknown but its log derivative is learned from the training data.
}
\label{fig:dm}
\end{center}
\end{figure}

As Fig.~\ref{fig:dm} demonstrates, in the limit of continuous time, the forward process, during which noise is added, follows a stochastic differential equation (SDE),
\begin{equation}
\frac{d \phi}{d\xi } = f(\phi, \xi) + g(\xi) \eta(\xi),
\label{eq:sde}
\end{equation}
where the Langevin time $\xi$ goes from $0$ to $T\equiv 1$, $\eta$ is the noise term independently added to all field degrees of freedom in the field configurations $\phi$, satisfying {$\langle \eta(\xi)\eta(\xi')\rangle= \delta(\xi-\xi')$}, and $f(\phi, \xi)$ is the vector drift term. The square of the scalar $g(\xi)$ is the time-dependent scalar \textit{diffusion coefficient}. In our case, the forward drift term $f(\phi,\xi)$ is set to zero and the variance-expanding scheme is chosen, see, e.g., Ref.~\cite{song2020score}, with $g(\xi) = \sigma^{\xi}$. This way, the transition kernel in the forward process  remains Gaussian, due to the additivity of Gaussian white noise,
\begin{equation}
    p_{\xi}(\phi_\xi | \phi_0) = \mathcal{N}\bigg(\phi_\xi; \phi_0, \sigma_\xi \mathbf{I}\bigg),
    \qquad
    \sigma_\xi = \frac{1}{\log \sigma^2}(\sigma^{2\xi} - 1).
    \label{eq:sbm}
\end{equation}
In this setup, we can add noise at any $0<\xi\leq T$, to achieve arbitrarily perturbed data without integration. In our implementation, we choose a geometric sequence of $\xi$ to be training points.

The reverse denoising process corresponding to Eq.~\eqref{eq:sde} is accomplished by a reverse SDE, representing an evolution backwards in time~\cite{anderson:1982reversetime},
\begin{equation}
  \frac{d\phi}{dt} = \left[ f(\phi, t) - g^2(t)\nabla_{\phi}\log p_t(\phi) \right] 
  + g(t) \bar{\eta}(t),
  \label{eq:rsde}
\end{equation}
where the reverse time $t$ runs from $T$ to $0$ and $\bar{\eta}$ is a noise term in the reverse time direction, still satisfying $\langle \bar \eta(t) \bar \eta(t')\rangle = \delta(t-t')$. Importantly, the drift term in Eq.~\eqref{eq:rsde} includes the gradient of the logarithm of $p_t(\phi)$, the probabilistic distribution at time $t$ in the forward diffusion process.

In Eqs. (3.1) and (3.3), the variables $\xi$ and $t$ are defined as continuous parameters in the interval $[0, 1]$ to describe the underlying diffusion process. In practical implementation, we evaluate these parameters at discrete noise levels $\xi_i$ following a geometric sequence (e.g., $\xi_i = \xi_0 \cdot r^i$). It is important to note that Eq. (3.2) defines the analytical marginal noise scale $\sigma_\xi$ for any given intensity $\xi$, rather than being derived from a path-dependent integration of a time-varying schedule. This definition allows us to sample $\xi$ independently from a continuous range during training while maintaining a consistent discrete schedule during the sampling phase.

\subsection{Noise Conditional Score Networks}
\label{sec:ncsn}

Typically, the gradients $\nabla_{\phi}\log p_t(\phi)$ in Eq.~\eqref{eq:rsde} are unknown. However, one can model a score function, $s(\phi, \xi)$, to match the gradients~\cite{song2020score}. Due to demands of the scalability and ability to efficiently handle high-dimensional parameter spaces, we employ a deep neural network to model the score function. To further improve sampling efficiency, we introduce a noise scheme for the forward process, in which $\xi_i$ ($i=1, \ldots, N_T$) is a positive geometric sequence that satisfies $\xi_1/\xi_2 = \cdots = \xi_{N_T-1}/\xi_{N_T} > 1$. 

Consequently, we aim to train a conditional neural network to estimate the score of perturbed field configurations distribution, i.e.,
\begin{equation}
\forall \xi \in \{ \xi_i \}_{i=1}^{N_T} : \quad s_{\theta}(\phi, \xi) \approx \nabla_{\phi_\xi} \log p_{\xi}(\phi_\xi  | \phi_0),
\end{equation}
as defined in Eq.~\eqref{eq:sbm}. Here $\theta$ denotes the parameters of the neural network. In training, we add known random noise to the data and train the network to estimate the noise scale. For a given $\xi$, the denoising \textit{score matching} objective is to minimize the loss function,
\begin{equation}
\ell(\theta; \xi) \triangleq \frac{1}{2} \mathbb{E}_{p_{\text{data}}(\phi)} \mathbb{E}_{\tilde{\phi} \sim \mathcal{N}(\phi, \sigma_\xi^2 I)} \left[ \left\lVert s_{\theta} (\tilde{\phi}, \xi) + \frac{\tilde{\phi} - \phi_0}{\sigma_\xi^2} \right\rVert_2^2 \right].
\end{equation}
This is combined for all $ \xi \in \{ \xi_i \}_{i=1}^{N_T} $ to get one unified objective,
\begin{equation}
\mathcal{L} (\theta; \{ \xi_i \}_{i=1}^{N_T}) \triangleq \frac{1}{N_T} \sum_{i=1}^{N_T} \lambda(\xi_i) \ell (\theta; \xi_i),
\end{equation}
where $ \lambda(\xi) = \sigma_\xi^2 $~\cite{song2020score}. Under this choice, we have
\begin{equation}
\lambda(\xi) \ell(\theta; \xi) = \frac{1}{2} \mathbb{E} \left[ \left\lVert \sigma_\xi s_{\theta}(\tilde{\phi}, \xi) + \frac{\tilde{\phi} - \phi_0}{\sigma_\xi} \right\rVert_2^2 \right].
\end{equation}
Since $(\tilde{\phi} - \phi_0)/\sigma_\xi \sim \mathcal{N}(0, I)$ and empirically~\cite{song2019generative} $\lVert \sigma_\xi s_{\theta}(\phi, \xi) \rVert_2 \propto 1 $, 
we can deduce that the order of magnitude of $ \lambda(\xi) \ell(\theta; \xi) $ is independent of $ \xi $. 

\begin{figure}[h]
\begin{center}
\includegraphics[width=0.85\textwidth]{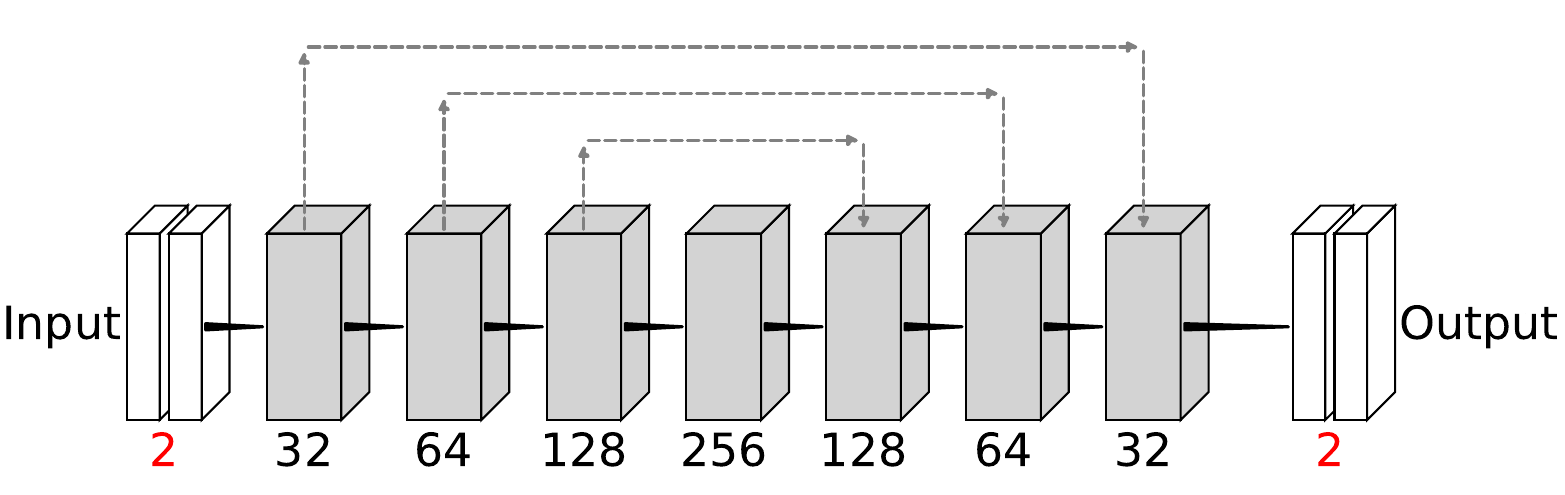}
\caption{Net architecture. The gray lines with arrows denote skip connections between the encoding and decoding paths. In this study, the default input shape is specified as $(\cdot,2,16,16)$, where the first dimension represents the batch size, the second corresponds to the channel dimension, and the last two denote the spatial dimensions of the field configuration.}\label{fig:unet}
\end{center}

\end{figure}

With regard to the specific network architecture, similar as in Ref.~\cite{Wang:2023exq}, we adopt a fully-convolutional architecture with skip connection as shown in Fig.~\ref{fig:unet} to model the score function. It employs an encoder-decoder structure with symmetric skip connections. The receptive field grows as the network deepens, but each convolution acts locally, making it adaptable to arbitrarily large input sizes.

\noindent\textbf{Input and Output.}
The input tensor is defined as $ X \in \mathbb{R}^{N\times2\times L\times L},$ where $N$ denotes the batch size, $2$ represents the two gauge link channels: angles of link variables in two dimensions, and $L$ (default $16$) is the spatial lattice size. The output tensor $Y$ has the same shape and approximates the score function, $ s_{\theta}(\phi, \xi).$ For sampling, the lattice size $L$ can be adjusted accordingly.

\el\noindent
\textbf{Fully-Convolutional Net Architecture.}
The encoder applies convolutional layers,
\begin{equation}
    h^{(l+1)} = \sigma\left(W^{(l)} * h^{(l)} + b^{(l)} + d^{(l)}(e(\xi))\right),
\end{equation}
where the noise (time) dependence is introduced via an embedding function of the noise parameter $\xi$. Specifically, we first generate a Gaussian random feature embedding of $\xi$,
\begin{equation}
    e(\xi) = [\sin(2\pi W\xi), \cos(2\pi W\xi)],
\end{equation}
where \(W\) in Eq.\,(3.9) is a shared “frequency embedding” matrix of shape  
\[
W \in \mathbb{R}^{d_{\mathrm{embed}} \times d_{\xi}},
\]  
where \(d_{\xi}\) is the dimension of the noise/time input (typically \(1\)) and \(d_{\mathrm{embed}}\) is the embedding dimension. This \(W\) is shared across all layers and spatial locations (i.e.\ it does not vary per lattice site).

The function \(d^{(l)}\) is implemented as a small fully-connected network,
\[
d^{(l)}: \mathbb{R}^{d_{\mathrm{embed}}} \to \mathbb{R}^{C_{\mathrm{out}}},
\]  
where \(C_{\mathrm{out}}\) is the number of feature channels. Its output is then broadcast spatially (i.e.\ replicated across all lattice sites) and added to the convolutional output at each spatial position.
Thus, noise dependence explicitly modulates intermediate feature representations at each stage of the fully convolutional network.

Each layer follows group normalization, $\hat{h}^{(l)}= {(h^{(l)}-\mu_G)}/{\sqrt{\sigma_G^2+\epsilon}}.$, where \(h^{(l)}\) denotes the input features at layer \(l\), \(\mu_G\) and \(\sigma_G\) are respectively the mean and standard deviation computed over the feature channels for each normalization group \(G\), and \(\epsilon\) is a small constant added for numerical stability. The decoder utilizes transposed convolutions,
\begin{equation}
h^{(l-1)}=\sigma\Big(W^{(l-1)}_T\circledast h^{(l)}+b^{(l-1)}\Big),
\end{equation}
where $\circledast$ denotes the deconvolution operation.
\begin{equation}
\tilde{h}^{(l-1)} = \operatorname{concat}\big(E^{(l-1)}, h^{(l-1)}\big),
\end{equation}
where $E^{(l-1)}$ indicates the corresponding encoder layer. Finally, the network output is normalized by the standard deviation of the corresponding Gaussian noise.

\el\noindent
\textbf{Locality and Periodic Padding}.
The lattice gauge action is constructed from Wilson loops, ensuring local interactions. Accordingly, the fully convolutional network (FCN) computes outputs locally, as 
\begin{equation}
Y_{i,j} = f\Big( \{X_{m,n} \mid m \in \mathcal{N}(i),\, n \in \mathcal{N}(j) \} \Big).
\end{equation}
It denotes the collection (or local patch) of input values that lie within a neighbourhood $\mathcal{N}(i)$ around index $i$ and $\mathcal{N}(j)$ around index $j$. Periodic boundary conditions are enforced by using periodic padding, meaning that if an index exceeds the lattice boundaries, it wraps around according to the modulo operation. We note that due to the intrinsic periodic boundary conditions of the lattice, our setup differs from the conventional U-Net used in diffusion models. In contrast to the standard U-Net, which progressively reduces and then recovers spatial resolution to combine features of different granularities via skip connections, we keep the lattice size fixed throughout the network. This design naturally respects the periodic structure of the gauge fields.

\section{Physics-Conditioned Sampler}
\label{sec:physics_conditioned_sampler}

Sampling in diffusion models can be realized \cite{Wang:2023exq} as a variant of stochastic quantization~\cite{Parisi:1980ys}, an approach for quantizing field theories that is particularly useful for gauge theories ~\cite{Damgaard:1987rr,Namiki:1993fd}. In stochastic quantization, one introduces a fictitious time $\tau$, extending a field $\phi(x)$ to $\phi(x,\tau)$, which then evolves according to a Langevin equation,
\begin{equation} 
\frac{\partial \phi(x, \tau)}{\partial \tau} = -\frac{\delta S[\phi]}{\delta \phi(x,\tau)} + \sqrt{2}\eta(x,\tau), \label{eq:tl}
\end{equation}
where $\eta(x,\tau)$ is a Gaussian white noise. After a sufficient number of Langevin steps, the system reaches equilibrium with a distribution $p(\phi) \propto e^{-S[\phi]}$. This formulation effectively transforms the path integral measure of the quantum field theory into a stationary distribution of a stochastic process.

In our case, diffusion models generate samples via a similar stochastic differential equation derived from the variance-expanding version of Eq.~\eqref{eq:rsde},
\begin{equation} 
\frac{\partial \phi(x, t)}{\partial t} = -g^2(t) s_{\hat{\theta}}(\phi(x,t),t) + g(t)\eta(x, t), 
\end{equation} 
starting from a Gaussian prior for $\phi(x)$. By gradually decreasing $t$ from $T=1$ to 0 (where the noise vanishes), the resulting sample distribution controlled by the score function becomes consistent with increasingly close to the target distribution used for training.

\subsection{Physics-Conditioned Score Function}

In this study, our target is to sample configurations at different values of the coupling $\beta$ using a diffusion model trained at a single specific coupling $\beta_0$, without requiring additional training. To achieve this, we construct a \textit{physics-conditioned sampler} by drawing upon principles from statistical field theory. Configurations are sampled from the physical distribution,
$p(\phi) = \exp[-S(\phi;\beta)]/Z$, where $\beta$ is a parameter (the inverse coupling in our case). In the pure gauge case, the action is separable as $S(\phi;\beta) = \beta \tilde{S}(\phi)$, and one can easily perform simulations at a reference value $\beta_0$ and then use Ferrenberg-Swendsen or histogram reweighting \cite{PhysRevLett.61.2635} to obtain results for other values of $\beta$, provided the sampled ensembles have sufficient overlap. This idea has been proposed and validated for other field systems with different generative models~\cite{Wang:2020nnsm,Xu:2024tjp}.

In diffusion models, the network-based score function $s_\theta$ is trained to approximate the drift term $\nabla_{\phi}\log p_t(\phi)$ in the reverse diffusion process. In contrast, in Langevin simulations of fields, the drift term is given by the functional derivative of the action, $-{\delta S[\phi]}/{\delta \phi(x,\tau)}$. If the diffusion model is well-trained, when the noise scale reaches $0$, the score function should approximate the actual drift term, $s_{\hat{\theta}}(\phi(x,0),0;\beta) \simeq -{\delta S(\phi;\beta)}/{\delta \phi(x)}$. For separable actions, such as in pure gauge theories, both the action and its derivative scale linearly with the coupling $\beta$. 
This suggests that a well-trained score network at a reference coupling $\beta_0$ can directly replace the drift term in Langevin simulations at arbitrary $\beta$ when equilibrium is reached around the final steps. Then one can generate configurations at different values of $\beta$ from the trained diffusion model directly, using the sampler to be introduced in the next section.

\subsection{Metropolis-Adjusted Annealed Langevin Algorithm}
\label{sec:mala}

\begin{figure}[hbtp!]
\begin{center}
    \includegraphics[width=0.85\textwidth]{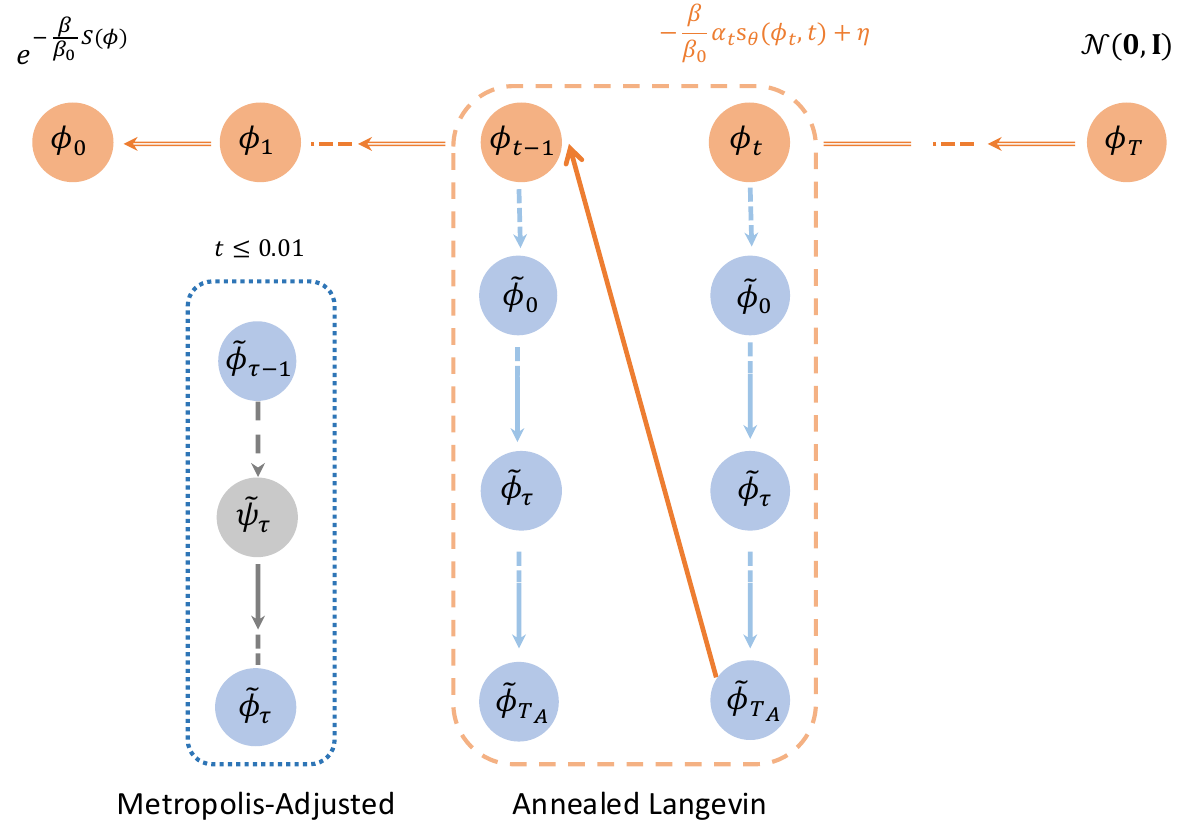}
\end{center}
\caption{\textbf{Metropolis-adjusted annealed Langevin sampler.} The evolution along the horizontal axis represents the diffusion model running backwards in time $t$ from the naive distribution $\mathcal{N}({\bf0, I})$ to the target physical distribution. The evolution along the vertical axis corresponds to the annealed Langevin dynamics with time $\tau$ and a fixed drift term, where the first dashed arrows indicate the copy operation. To distinguish them, we use a different notation to label the field variables.}
\label{fig:mala}
\end{figure}

As explained earlier, generating samples for U(1) lattice gauge theory at different coupling values is only feasible if equilibrium is achieved in the final noise-scale steps. To ensure this, we use the \textit{annealed Langevin scheme} proposed in Ref.~\cite{song2019generative}. In Fig.~\ref{fig:mala}, the reverse denoising process is demonstrated along the horizontal time axis, with different noise-scales at different $t$. Each orange double arrow can be expanded into a dashed orange box describing the annealed scheme. This is where we run evolution along another time direction, $0<\tau\leq T_A$, with the fixed drift term $-\frac{\beta}{\beta_0}g^2(t) s_{\hat{\theta}}(\phi_t, t)$. The final state of the annealed Langevin evolution is copied as the initial state at the next noise scale. By performing multiple Langevin updates at each progressively decreasing noise scale $t_i$, one ensures that the distribution approaches $p(\phi,t_i) \propto e^{-S_{DM}(\phi,t_i)}$, where $\partial S_{DM}(\phi,t_i)/\partial\phi = -\frac{\beta}{\beta_0} \hat{s}_\theta(\phi,t_i)$. This will eventually bring us to $S_{DM}(\phi,t_i \rightarrow 0) \simeq S(\phi;\beta)$ in the final steps. The detailed set-up of the annealed scheme can be found in Appendix~\ref{app:langevin}.

In addition, to ensure that a Markov chain has the correct stationary measure, one can add a Metropolis-Hastings filter, i.e., an accept-reject step, along it. We extended the Metropolis-adjusted Langevin algorithm (MALA)~\cite{roberts1998optimal,girolami2011riemann}  to Metropolis-Adjusted Annealed Langevin algorithm (MAALA) by building multi-Markov chains with the annealed process to update configurations and adjusting the chains by Metropolis updates  during the final steps, $0\leq t \leq 0.001$.\footnote{In our case, $t\leq 0.001$ means $N_C= 35$ discrete time steps along the diffusion model backward direction in the set-up of noising scheme, using $0.9^{N_T-N_C}\leq 0.001$. More details can be found in Appendix~\ref{app:langevin}.} This will prepare $N_C$ chains for collecting configurations to estimate observables accurately. This sampler will be referred as DM-MAALA (DM for short in the following context). As shown in the dashed blue box of Fig.~\ref{fig:mala}, the Metropolis-adjusted procedure is employed in the annealed process for the last $N_C$ noise-scales. Since $S_{DM}(\phi,t_i \rightarrow 0) \simeq S(\phi;\beta)$, the sampler can reach a reasonably high acceptance rate as shown in Appendix~\ref{app:acc}.

In a general MALA~\cite{roberts1998optimal,girolami2011riemann}, from the current state $\phi_\tau$, the proposed update is
\begin{equation}
    \psi_{\tau+1} = \phi_\tau + \alpha_i F(\phi_\tau,\tau) + \sqrt{2\alpha_i} \eta, 
\end{equation}
where $\alpha_i $ is a redefined noise-scale parameter which can be found in Appendix~\ref{app:langevin}, and $F$ is one proposal kernel, which is the score function in DM. The accept-reject step is
\begin{equation}
\phi_{\tau+1} =
    \begin{cases} 
    \psi_{\tau+1} & \text{with probability } \min\left\{1, \frac{p(\psi_{\tau+1}) q(\phi_\tau|\psi_{\tau+1})}{p(\phi_\tau) q(\psi_{\tau+1}|\phi_\tau)} \right\}, \\
    \phi_\tau & \text{with the remaining probability,}
    \end{cases}
\end{equation}
where $p(\phi)$ is computed from the known physical distribution. Note that conditioned on $\phi_\tau$, the proposal step boils down to drawing a Gaussian with mean $ \phi_\tau + \alpha_i F(\phi_{\tau},\tau)$ and variance $2\alpha_i I_\tau$. Hence, the proposal kernel has the explicit form
\begin{equation}
q(\phi_\tau|\psi_{\tau+1}) = \frac{1}{(4 \pi\alpha_i)^{n/2}} \exp\left(-\frac{1}{4\alpha_i}
\left\|\phi_\tau - (\psi_{\tau+1} +  \alpha_i F(\psi_{\tau+1},\tau+1))\right\|_2^2\right),
\end{equation}
where $n$ is the number of degrees of freedom, which is $n=2V$ in our case. The transition probability can then be computed explicitly, as

\begin{align}
\frac{p(\psi_{\tau+1}) q(\phi_\tau|\psi_{\tau+1})}{p(\phi_\tau) q(\psi_{\tau+1}|{\phi_\tau})} = 
\exp\bigg(
-S(\psi_{\tau+1}) - \frac{1}{4\alpha_i} \|\phi_\tau - (\psi_{\tau+1} +\alpha_i F(\psi_{\tau+1},\tau+1))\|_2^2 & \nonumber\\
+ S(\phi_\tau) + \frac{1}{4\alpha_i} \|\psi_{\tau+1} - (\phi_{\tau} + \alpha_i F(\phi_{\tau},\tau))\|_2^2
\bigg). &
\end{align}

Combined with the annealed scheme, MAALA can generate exact configurations at different inverse couplings $\beta$. A detailed set-up can be found in Algorithm~\ref{alg:MALAannealed-langevin} of Appendix~\ref{app:langevin}. In fact, the classical MALA simulation is obtained by replacing the time-dependent score function by the time-independent physical drift term derived from the given action. We will compare our results with the MALA simulation according to Eq.~(\ref{eq:tl}). In this study, the results of the diffusion model are all generated by the Metropolis-adjusted Langevin annealed algorithm.

\section{Numerical Experiments}
\label{sec:numerical_experiments}

For building a training data set, we generated 30720 configurations at $\beta = 1.0$ on a $16\times 16$ lattice. For testing, we generated 1024 configurations at $\beta = 1.0, 3.0, 5.0, 7.0, 9.0, 11.0$ on different lattices of size $8\times 8$, $16\times 16$ and $32\times 32$. The gauge symmetry of the theory is employed for data augmentation, where a set of configurations resulting from a gauge transformation is treated as a new set of configurations. This informs our neural network that, despite numerical differences, the underlying physical properties remain the same after gauge transformations. 

In this work, configurations for training are generated by HMC
on one single chain 
, which is an efficient method for generating gauge field configurations, at least at $\beta = 1.0$. To generate better samples, for each Markov chain in HMC, we have a ``200-step burn-in'' steps\footnote{``Burn-in'' in this study refers to the generation steps before Metropolis is employed.} for thermalization, and after 128 additional steps, we select the final step of the 128-step update as a configuration. Here one step indicates one MC trajectory with a well-established HMC integrator, as in Ref.~\cite{Wang:2023exq}. 
%

A comprehensive comparison between the algorithms discussed here can be found in Appendix~\ref{app:para}.

\subsection{Performance in the Unfrozen Region}

\subsubsection{Training and validation}

\begin{figure}[hbtp!]
\begin{center}
    \includegraphics[width=0.45\textwidth]{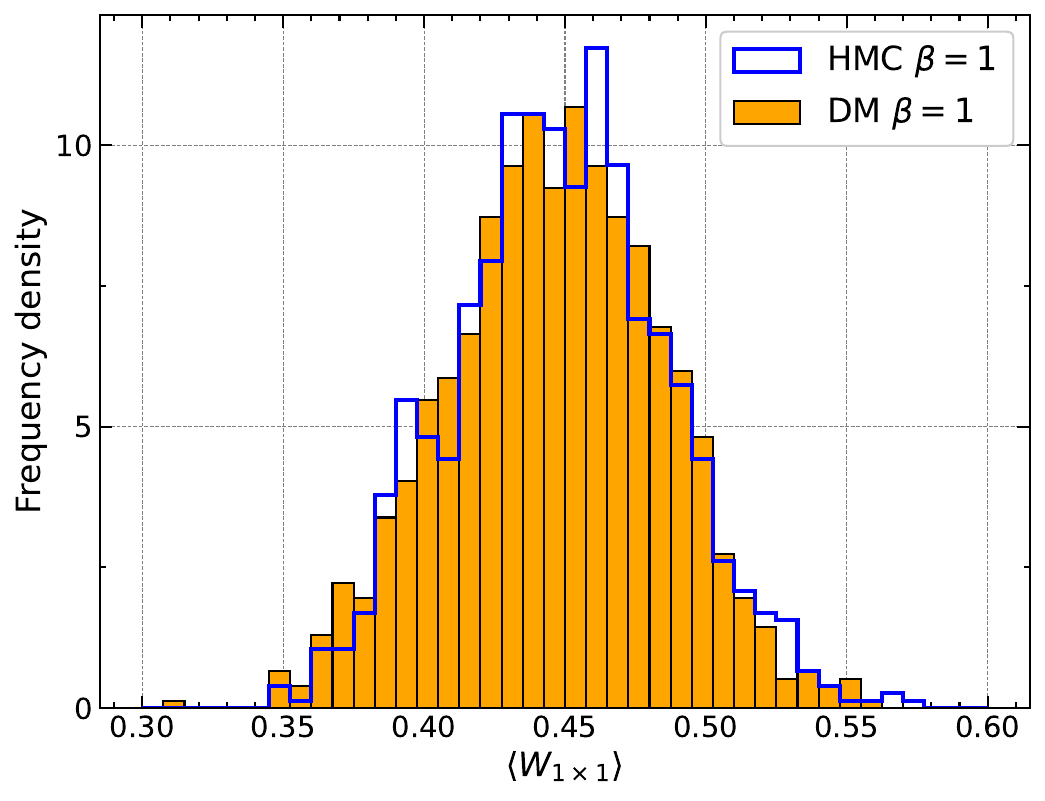}
    \includegraphics[width=0.462\textwidth]{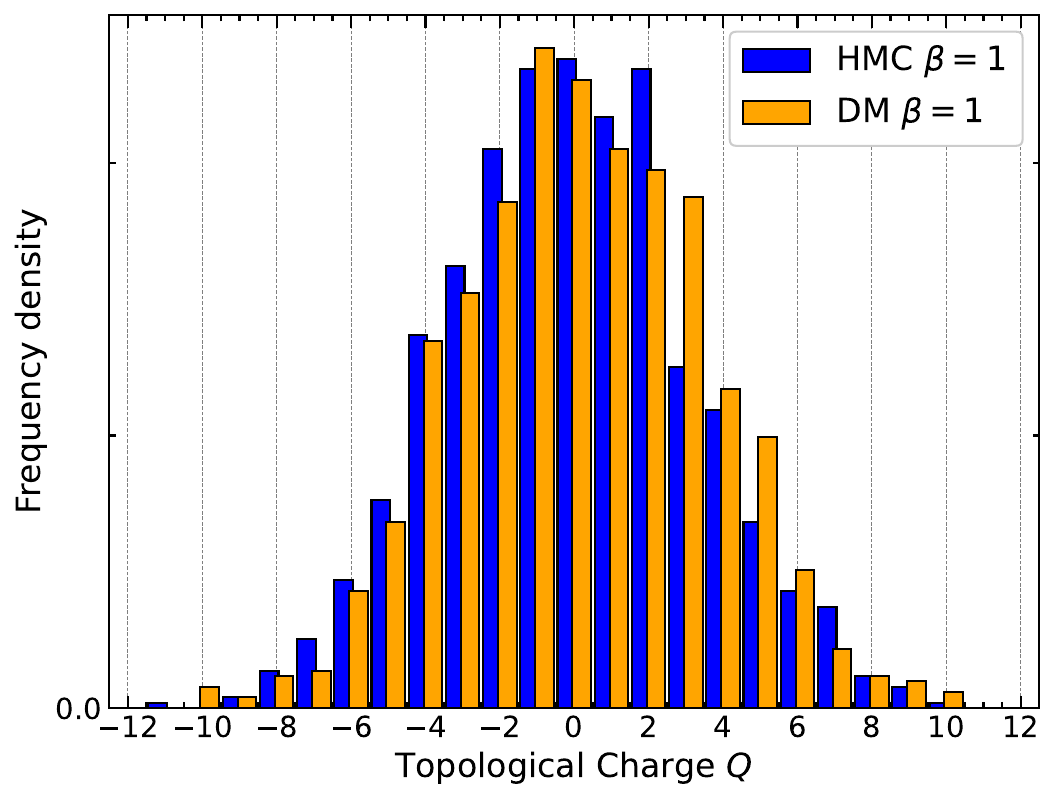}
\end{center}
\caption{Histograms of the Wilson loop (left)  and the topological charge (right) at $\beta = 1, L=16$ from the testing data-set (HMC) and from the trained diffusion model (DM).}
    \label{fig:beta1_L16}
\end{figure}

First, we evaluate the training performance in the topologically unfrozen region at $\beta = 1$. All configurations are generated by the original sampler, as introduced in Ref.~\cite{Wang:2023exq}, from the trained diffusion model. In Fig.~\ref{fig:beta1_L16}, the distributions of the Wilson loop (left) and the topological charge (right) are shown as orange bins, demonstrating consistency with the HMC results, which are represented by blue envelopes and bins. Qualitatively, the diffusion model achieves comparable performance to HMC in estimating both local and topological observables in the topologically unfrozen region. More quantitative results are shown in the following sections with our new sampler.

\subsubsection{Generation at different lattice sizes}

As a further validation and direct extrapolation, the performance across different lattice sizes is evaluated, see Table~\ref{tab:L_beta1}. These results demonstrate that the estimates of the Wilson loop (left) remain consistent across all lattice sizes, agreeing with the analytical result, see Appendix~\ref{app:exact}. Similarly, the calculation of the topological charge (right) in Table~\ref{tab:L_beta1} exhibits stability, indicating that the DM accurately captures the underlying physics.
The number of independent configurations is 1024 in all cases.
Since the DM was trained on a $16\times 16$ lattice only, 
these findings highlight the capability of DMs to train on small volumes and generalize to larger volumes without loss of accuracy, owing to their fully convolutional architecture introduced in Section~\ref{sec:ncsn}. This property significantly reduces memory requirements during training, making DMs a feasible approach for directly generating  configurations in larger systems.


\begin{table}[hbpt!] 
    \centering 
    \caption{Comparison of observables for $\beta=1$ at different lattice sizes}
    \label{tab:L_beta1}
    \renewcommand{\arraystretch}{1.3} 
    \resizebox{\textwidth}{!}{
    \begin{tabular}{|c|c c c c|c c c c|}
        \hline
        \multirow{2}{*}{Lattice Size ($L$)} & \multicolumn{4}{c|}{$1 \times 1$ Wilson Loop} & \multicolumn{4}{c|}{Topological Susceptibility} \\
        \cline{2-9}
        & HMC & DM & MALA & Exact & HMC & DM & MALA & Exact \\
        \hline
        8  & 0.447(72)  & 0.445(74)  & 0.443(80)  & 0.446  & 0.0402(17)  & 0.0413(18)  & 0.0418(18)  & 0.0406  \\
        16 & 0.447(37)  & 0.446(37)  & 0.444(36)  & 0.446  & 0.0416(16)  & 0.0422(17)  & 0.0421(20)  & 0.0406  \\
        32 & 0.446(18)  & 0.445(19)  & 0.445(18)  & 0.446  & 0.0428(19)  & 0.0415(18)  & 0.0412(17)  & 0.0406  \\
        64 & 0.446(9)   & 0.446(11)  & 0.445(9)   & 0.446  & 0.0426(19)  & 0.0427(20)  & 0.0420(19)  & 0.0406  \\
        \hline
    \end{tabular}
    }
\end{table}

\subsection{Generation in the Frozen Region}

\subsubsection{Mitigate freezing}

Next, we evaluate the model’s performance at different coupling values, without additional training, in the topologically frozen region using the physics-conditioned sampler introduced in Section~\ref{sec:mala}.

\begin{figure}[hbtp!]
\begin{center}
    \includegraphics[width=0.45\textwidth]{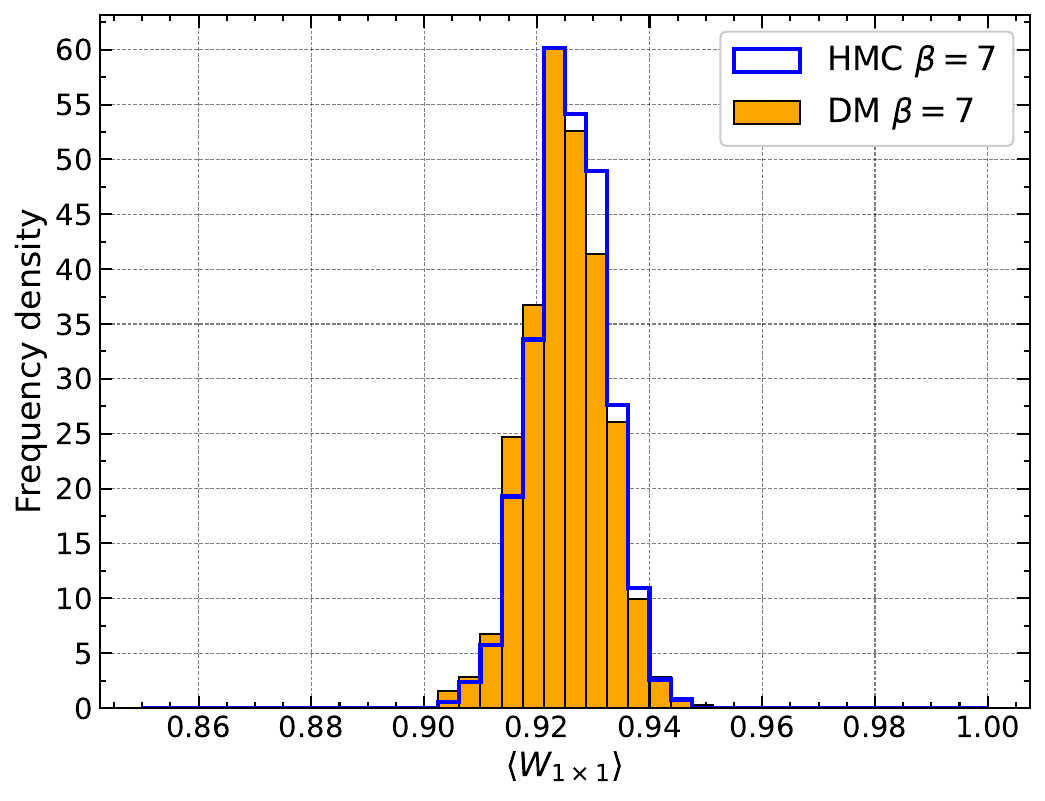}
    \includegraphics[width=0.453\textwidth]{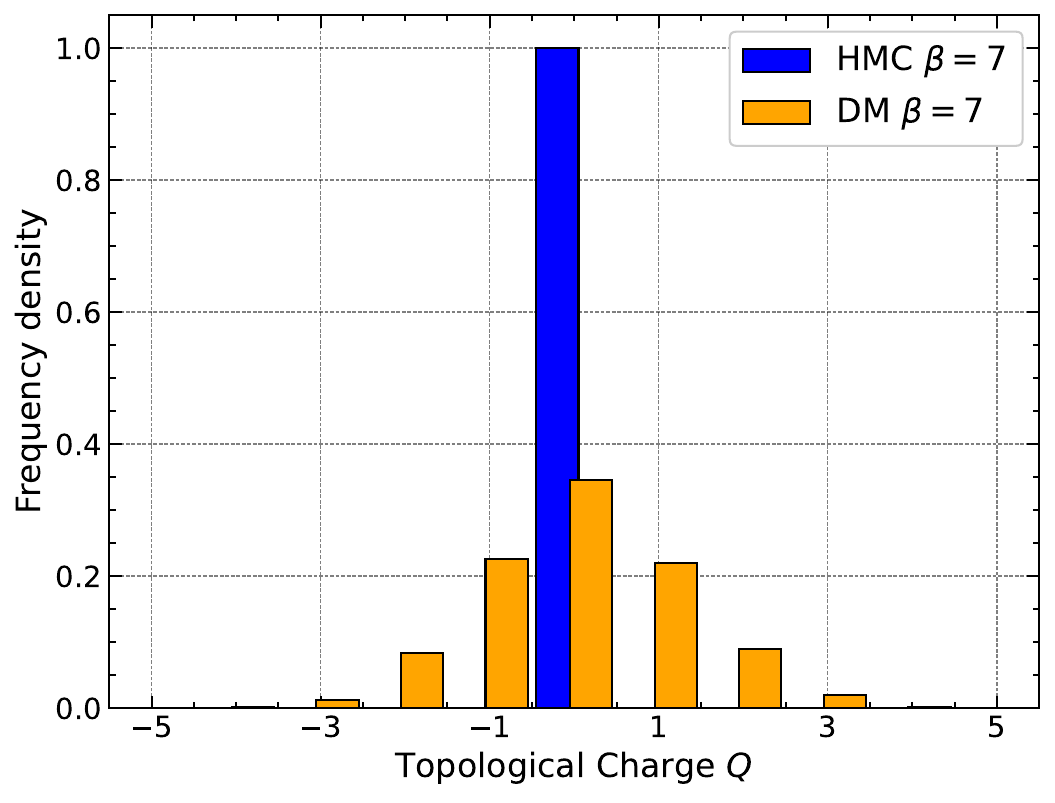}
\end{center}
\caption{
    Comparison of distributions for the Wilson loop (left)  and the topological charge (right) at $\beta = 7$ from the testing data-set (HMC) and from the DM trained at $\beta=1$ but conditioned at $\beta=7$. The number of independent configurations is 1024 in both cases.
}
\label{fig:beta-7}
\end{figure}

\begin{table}[htbp!] 
    \centering 
    \caption{Comparison of the $l \times l$ Wilson Loops for $L=16, \beta=7$} 
    \label{tab:L16_beta7} 
    \renewcommand{\arraystretch}{1.2}
    \setlength{\tabcolsep}{12pt}
    \begin{tabular}{|c|cccc|}
        \hline
        Loop size ($l$) & \multicolumn{1}{c}{HMC} & \multicolumn{1}{c}{DM} & \multicolumn{1}{c}{MALA} & \multicolumn{1}{c|}{Exact} \\ 
        \hline
        1 & 0.926(7)  & 0.926(7)  & 0.924(6)  & 0.926  \\ 
        2 & 0.737(31) & 0.737(32) & 0.730(34) & 0.734  \\ 
        3 & 0.510(67) & 0.496(72) & 0.489(73) & 0.498  \\ 
        4 & 0.311(97) & 0.283(96) & 0.283(106) & 0.290  \\
        \hline
    \end{tabular}
\end{table}

Using the same trained diffusion model, we generate 1024 configurations at $\beta=7$ and compare them with a set of 1024 configurations generated via HMC. The results are presented in Fig.~\ref{fig:beta-7}. We observe that the Wilson loop distributions (left) remain consistent between the two approaches. A more detailed comparison of Wilson loop at different sizes is provided in Table~\ref{tab:L16_beta7}, which also includes a comparison with MALA. 

However, for the topological charge in Fig.~\ref{fig:beta-7} (right), HMC exhibits topological freezing, sampling only values of the topological charge $Q$ near or at zero
 -- see Appendix~\ref{app:para} for a comparison between all the algorithms discussed above. 
%
%
In contrast, 
the physics-conditioned sampler using the DM explores a broader range of topological sectors, leading to a larger topological susceptibility that is consistent with the analytical results.
In addition, one can also estimate the likelihood for different conditions to evaluate whether the model is approaching the target physics distribution, which is discussed in Appendix~\ref{app:likelihood_estimation}.

\subsubsection{Generation at different lattice sizes and couplings}

Next we estimate observables at both different couplings and lattice sizes. Table~\ref{tab:L_beta7} presents a comparison of observables at fixed coupling $\beta = 7$ for lattice sizes $L=8\times 8, 16\times 16$ and $32\times 32$. The observables studied are the $1 \times 1$ Wilson loop and the topological susceptibility. For the three samplers, the Wilson loop values remain relatively stable across different lattice sizes and numerical methods, with all deviations within statistical uncertainties. However, the topological susceptibility shows significant variation depending on the method used. In particular, MALA yields notably higher susceptibility values compared to the exact results, while the HMC results indicate freezing. The DM remains closer to the exact predictions.

\begin{table}[hbpt!] 
    \centering 
    \caption{Comparison of observables for $\beta=7$ at different lattice sizes}
    \label{tab:L_beta7}
    \renewcommand{\arraystretch}{1.3}
    \resizebox{\textwidth}{!}{
    \begin{tabular}{|c|c c c c|c c c c|}
        \hline
        \multirow{2}{*}{Lattice Size ($L$)} & \multicolumn{4}{c|}{$1 \times 1$ Wilson Loop} & \multicolumn{4}{c|}{Topological Susceptibility} \\
        \cline{2-9}
        & HMC & DM & MALA & Exact & HMC & DM &  MALA & Exact \\
        \hline
        8  & 0.927(13)  & 0.926(13)  & 0.921(13)  & 0.926  & 0.00006(3)  & 0.0057(2)  & 0.0143(5)  & 0.0040  \\
        16 & 0.926(7)   & 0.926(7)   & 0.924(6)   & 0.926  & 0.00013(2)  & 0.0063(3)   & 0.0131(5)  & 0.0039  \\
        32 & 0.926(3)   & 0.925(4)   & 0.924(4)   & 0.926  & 0.00013(2)  & 0.0051(3)   & 0.0137(6)  & 0.0039  \\
        \hline
    \end{tabular}
    }
\end{table}

\begin{table}[hbpt!] 
    \centering 
    \caption{Comparison of observables for $L=16$ at different couplings}
    \label{tab:L_16}
    \renewcommand{\arraystretch}{1.3}
    \resizebox{\textwidth}{!}{
    \begin{tabular}{|c|c c c c|c c c c|}
        \hline
        \multirow{2}{*}{coupling ($\beta$)} & \multicolumn{4}{c|}{$1 \times 1$ Wilson Loop} & \multicolumn{4}{c|}{Topological Susceptibility} \\
        \cline{2-9}
        & HMC & DM & MALA & Exact & HMC & DM & MALA & Exact \\
        \hline
        3 & 0.811(17)   & 0.811(17)   & 0.809(17)   & 0.810  & 0.0096(4)  & 0.0114(6)   & 0.0106(14)  & 0.0111  \\
        5 & 0.894(9)   & 0.894(9)   & 0.891(10)   & 0.894  & 0.0048(2)  & 0.0058(3)   & 0.0075(3)  & 0.0057  \\
        7 & 0.926(7)   & 0.926(7)   & 0.924(6)   & 0.926  & 0.00013(2)  & 0.0063(3)   & 0.0131(5)  & 0.0039  \\
        9 & 0.944(3)   & 0.942(4)   & 0.940(6)   & 0.942  & 0  & 0.0053(3)   & 0.0154(7)  & 0.0029  \\
        11 & 0.954(3)   & 0.953(4)   & 0.950(5)   & 0.953  & 0  & 0.0058(3)   & 0.0165(13)  & 0.0024  \\
        \hline
    \end{tabular}
    }
\end{table}

Table~\ref{tab:L_16} examines the dependence of observables on the coupling constant, at a fixed lattice size of $16\times 16$. As $\beta$ increases, the Wilson loop approaches $1$, with reduced fluctuations. Similarly, the topological susceptibility decreases with increasing $\beta$, indicating a suppression of topological fluctuations at weak coupling.

We found that for the $1\times1$ Wilson loop, a DM trained at $\beta=1$ yields results consistent with the exact values when employed at larger beta values (we went up to $\beta=11$), underscoring the effectiveness of physics-conditioned diffusion models. For the topological susceptibility on the other hand, such an extrapolation was found to break down after $\beta \gtrsim 5$, leaving a more limited range of couplings available. Despite best efforts, this issue is currently an open problem.

\section{Summary and Outlook}
\label{sec:outlook}

In this study, we applied diffusion models to generate gauge field configurations. The correspondence between the stochastic differential equations encountered in diffusion models and in stochastic quantization enables the model to generate configurations at different lattice sizes and coupling constants without additional retraining. To ensure that the sampled distribution corresponds to the physical distribution, we combined the Metropolis-adjusted Langevin algorithm with the annealed Langevin sampler. Our results demonstrate that this multi-chain sampler effectively mitigates topological freezing at large $\beta$, and that the diffusion model proposes improved configurations for the Markov chains, leading to estimates of observables that remains consistent withe exact results when extrapolation is in intermediate interval.

Although DM-MAALA shows good ergodicity across topological sectors in early sampling stages~\cite{Zhu:2024kiu}, the DM generation cannot completely avoid the bias of topology distribution in the training set. Intuitively speaking, when extrapolation to larger $\beta$ values, only the magnitude of the learned score is changed, while the direction is unaffected. This may cause the generated configurations to remain biased towards the topology distribution of the training set. Going forward, it is crucial to investigate this further (and resolve it). To mitigate this, one possibility is to adopt \emph{topological‐sector–fixed sampling schemes}~\cite{Fukaya:2006ca} as current applications in HMC methods, and later merge or stitch configurations across sectors. Another practical improvement is to restrict extrapolation to \emph{small $\beta$ intervals} so that sectors overlap sufficiently and the sampler can jump among them easily in the Metropolis procedure. Furthermore, it is also possible to introduce \emph{auxiliary fields} or coupling variables to sample $Q$ directly~\cite{Chen:2024ddr}. Although it will introduce additional cost to the sampling procedure, it may offer a promising solution in our framework.

Apart from this difficulty, extending this approach to non-Abelian gauge theories, including fermionic systems, appears straightforward; however, the non-locality of fermion determinants in QCD-like theories remains a significant challenge as it may need a novel structure of network architecture to catch long-range correlations. The first attempt for a complex action (sign) problem has been reported recently~\cite{Habibi:2024fbn}. For sampling, improvements of annealed Langevin dynamics \cite{Albergo:2024trn} should be considered. Concerning the neural network architecture, the integration of diffusion models with neural networks specifically designed for lattice configurations~\cite{Favoni:2020reg,Nagai:2021bhh,Nagai:2025rok,Aarts:2025gyp} is a promising direction. The scalability of diffusion models suggests its potential for generating configurations on larger lattices at large $\beta$, requiring training only on samples from smaller lattice sizes and lower $\beta$ values. Our objective for the future is to develop a complementary approach that enhances the efficiency and accuracy of lattice Quantum Chromodynamics computations, particularly in regimes where traditional techniques face limitations.

\el \noindent
{\bf Notes} --
A related study has been released recently~\cite{Ranner:2024qtv}.

\acknowledgments

We thank Shiyang Chen and Diaa E. Habibi for comments and Drs.\ Akinori Tanaka, Tetsuo Hatsuda, Andreas Ipp, Gurtej Kanwar, Fernando Romero-L\'opez, and Shuzhe Shi for helpful discussions.
We thank the DEEP-IN working group at RIKEN-iTHEMS for support in the preparation of this paper.
WW and QZ are supported in part by Natural Science Foundation of China under grants No.\ 12125503 and 12335003, and SJTU Kunpeng \& Ascend Center of Excellence. GA is supported by STFC Consolidated Grant 
ST/X000648/1. KZ is supported by the CUHK-Shenzhen university development fund under grant No.\ UDF01003041 and UDF03003041, Ministry of Science and Technology of China under Grant No. 2024YFA1611004, NSFC fund under No. 92570117 and Shenzhen Peacock fund under No.\ 2023TC0179. L.~Wang is also supported by the RIKEN-TRIP initiative (RIKEN-Quantum), JSPS KAKENHI Grant No. 25H01560, and JST-BOOST Grant No.JPMJBY24H9.

\el\noindent
{\bf Research Data and Code Access} --
The code and data used for this manuscript can be found at \url{https://github.com/zzzqt/DM4U1}.

\el\noindent
{\bf Open Access Statement} -- For the purpose of open access, the authors have applied a Creative Commons Attribution (CC BY) licence to any Author Accepted Manuscript version arising.

\appendix

\section{Langevin Sampler}
\label{app:langevin}

\begin{algorithm}[hbtp!]
\caption{Annealed Langevin Sampler}
\label{alg:annealed-langevin}
\begin{algorithmic}[1] 
    \Require $\{t_i\}_{i=1}^{N_T}, \epsilon, N_A$
    \State Initialize ${\mathbf{\phi}}_0$
    \For {$i \leftarrow 1$ to $N_T$}
        \State $\alpha_i \leftarrow \epsilon \cdot g \cdot t_i^2 / t_{N_T}^2$ \Comment{$\alpha_i$ is the step size.}
        \State {$\tilde{\mathbf{\phi}}_0 \leftarrow {\mathbf{\phi}_{i-1}}$}
        \For {$\tau \leftarrow 1$ to $N_A$}
            \State Draw $\eta \sim \mathcal{N}(0, I)$
            \State $\tilde{\mathbf{\phi}}_\tau \leftarrow \tilde{\mathbf{\phi}}_{\tau-1} + \frac{\beta}{\beta_0} \alpha_i s_\theta(\tilde{\mathbf{\phi}}_{\tau-1}, t_i) + \sqrt{2\alpha_i} \eta$
        \EndFor
        \State ${\mathbf{\phi}}_{i} \leftarrow \tilde{\mathbf{\phi}}_{N_A}$
    \EndFor
    \State \textbf{return} ${\mathbf{\phi}}_{N_T}$
\end{algorithmic}
\end{algorithm}

\begin{algorithm}[hbtp!]
\caption{Metropolis-Adjusted Annealed Langevin Sampler}
\label{alg:MALAannealed-langevin}
\begin{algorithmic}[1] 
    \Require $\{t_i\}_{i=1}^{N_T}, \epsilon, N_A$
    \State Initialize ${\mathbf{\phi}}_0$
    \For {$i \leftarrow 1$ to $N_T$}
        \State $\alpha_i \leftarrow \epsilon \cdot g \cdot t_i^2 / t_{N_T}^2$ \Comment{$\alpha_i$ is the step size.}
        \State {$\tilde{\mathbf{\phi}}_0 \leftarrow {\mathbf{\phi}_{i-1}}$}
        \For {$\tau \leftarrow 1$ to $N_A$}
            \State Draw $\eta \sim \mathcal{N}(0, I)$
            \State Propose $\tilde{\mathbf{\psi}}_\tau = \tilde{\mathbf{\phi}}_{\tau-1} + \frac{\beta}{\beta_0} \alpha_i s_\theta(\tilde{\mathbf{\phi}}_{\tau-1}, t_i) + \sqrt{2\alpha_i} \eta$
            \If{$t_i < 0.001$} 
                \State $P_i \leftarrow \min\left\{1, \frac{p(\tilde{\psi}_{\tau}) q(\tilde{\phi}_{\tau-1}|\tilde{\psi}_{\tau})}{p(\tilde{\phi}_{\tau-1}) q(\tilde{\psi}_{\tau}|{\tilde{\phi}_{\tau-1}})} \right\}$
                \State Draw $k \sim \mathrm{Uniform}(0,1)$
                \If{$k_i < P_{i}$}
                    \State$\tilde{\mathbf{\phi}}_\tau \leftarrow \tilde{\mathbf{\psi}}_{\tau}$
                \Else
                    \State $\tilde{\mathbf{\phi}}_\tau \leftarrow \tilde{\mathbf{\phi}}_{\tau-1}$
                \EndIf
            \Else
                \State$\tilde{\mathbf{\phi}}_\tau \leftarrow \tilde{\mathbf{\psi}}_{\tau}$
            \EndIf
        \EndFor
        \State ${\mathbf{\phi}}_{i} \leftarrow \tilde{\mathbf{\phi}}_{N_A}$
    \EndFor
    \State \textbf{return} ${\mathbf{\phi}}_{N_T}$
\end{algorithmic}
\end{algorithm}
Algorithm~\ref{alg:annealed-langevin} provides the pseudocode for our annealed Langevin scheme. Here $g$ is the diffusion coefficient, ${t_i}$ is chosen as a geometric series starting from 1 with a common ratio of 0.9 and a total of $N_T=100$ steps, $\epsilon$ is fixed to $5e^{-9}$ such that $\epsilon \cdot t_0^2 / t_{N_T}^2 \sim 1$, and we set $N_A=1000$. The initial distribution is taken to be the standard normal distribution. Algorithm~\ref{alg:MALAannealed-langevin} is the pseudocode of the integration of the Metropolis-adjusted Langevin algorithm with the annealed scheme. If no further explanation is given, all related settings are the same as for the Algorithm~\ref{alg:annealed-langevin}. The transition probability is computed from the given physical action and the learned score functions.

\section{Autocorrelation Time}
\label{app:acc}

Although in our approach multiple Markov chains are employed to generate independent samples, the dynamics in a single chain can be compared with other Monte Carlo methods. For traditional Monte Carlo methods, the characteristic length of autocorrelations for any observable $\mathcal{O}$ can be defined by the integrated autocorrelation time $\tau^\mathrm{int}_\mathcal{O}$~\cite{Madras:1988ei}.

\begin{figure}[hbpt!]
\centering
\includegraphics[width=0.49\textwidth]{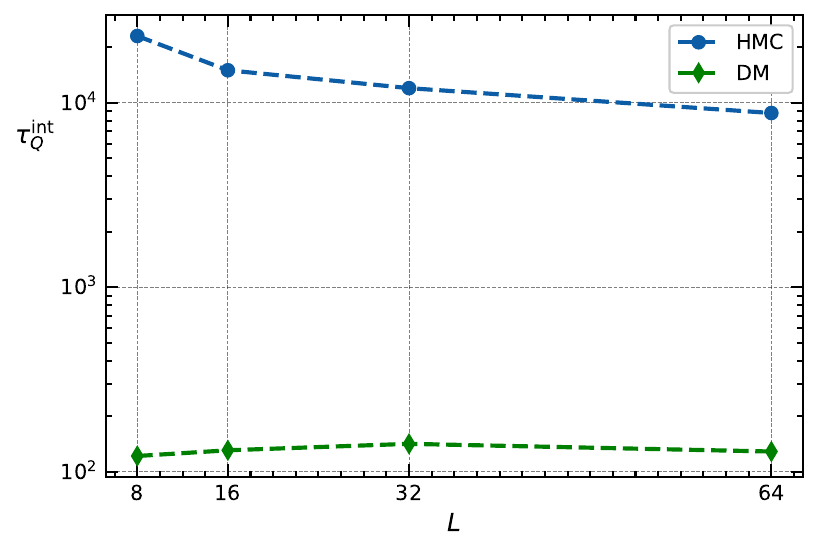}
\includegraphics[width=0.49\textwidth]{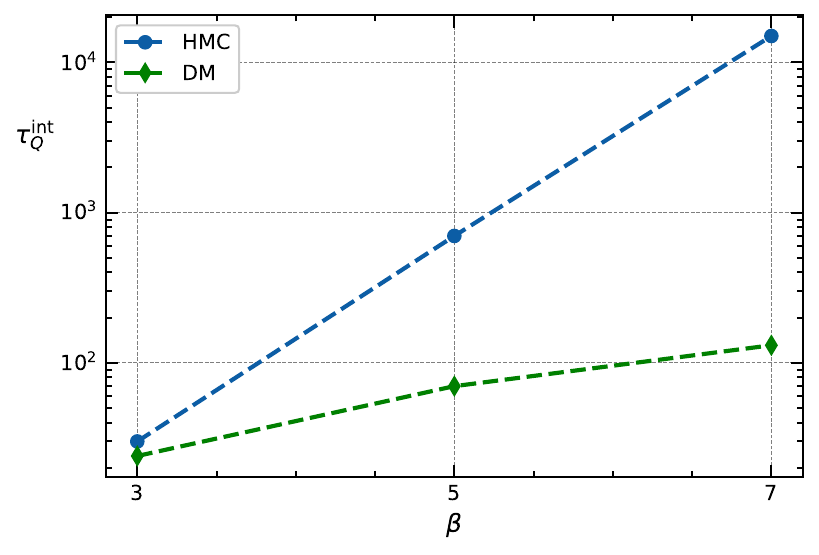}
\caption{
    Comparison of integrated autocorrelation time in a single Markov chain for HMC and the DM at $\beta=7$ for different lattice sizes $L$ (left), and at $L=16$ for different couplings $\beta$ (right) 
}
\label{fig:corr}
\end{figure}

\begin{figure}[h]
\centering
\includegraphics[width=0.49\textwidth]{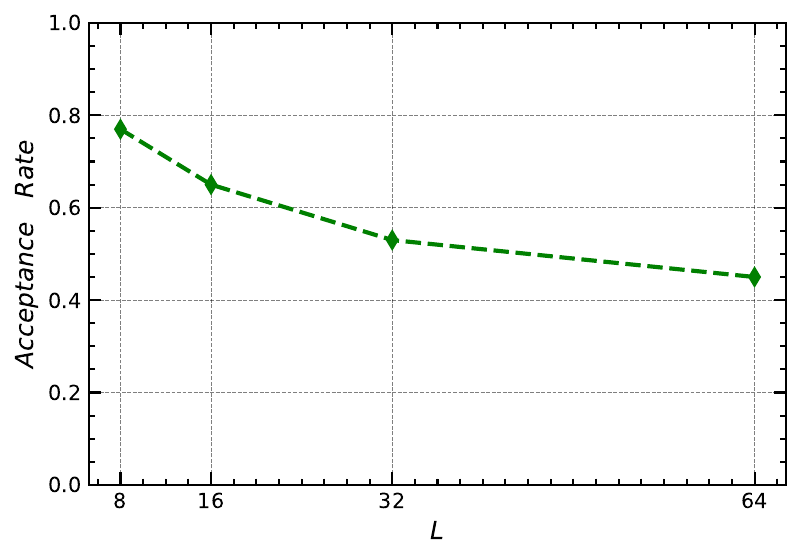}
\includegraphics[width=0.49\textwidth]{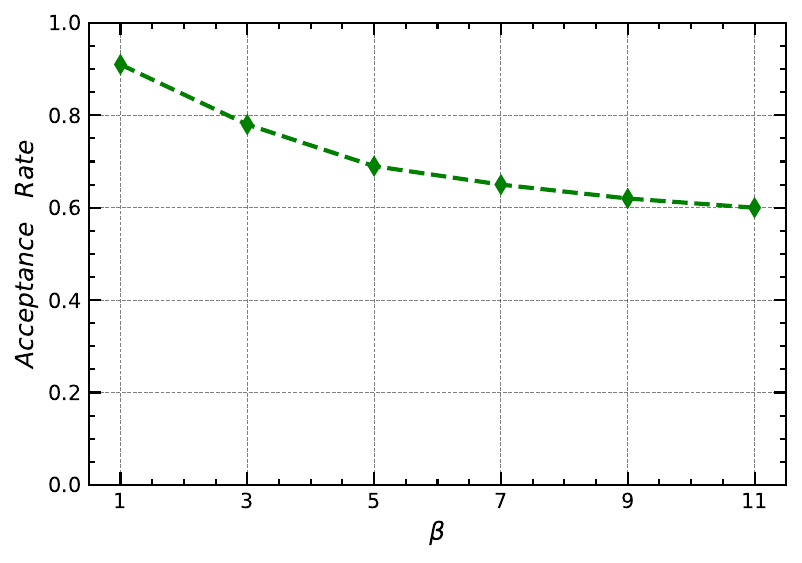}
\caption{
    Dependence of acceptance rate in a single Markov chain of a DM at $\beta=7$ for different lattice sizes $L$ (left), and at $L=16$ for different couplings $\beta$ (right) 
}
\label{fig:acc}
\end{figure}

Fig.~\ref{fig:corr} compares the integrated autocorrelation time of the topological charge, $\tau^\mathrm{int}_Q$, for HMC and DM. For both methods, $\tau^\mathrm{int}_Q$ increases as $\beta$ is raised. In a single Markov chain within HMC, once equilibrium is reached, the topological charge rarely evolves at large $\beta$, whereas the diffusion model exhibits significantly better ergodicity. DM also demonstrates stability across different lattice sizes, with $\tau^\mathrm{int}_Q \approx 130$ for all sizes at $\beta=7$. Additionally, we test a Metropolis-adjusted traditional Langevin sampler by replacing only the drift term in Algorithm~\ref{alg:MALAannealed-langevin} with $ -\delta S[\phi]/\delta \phi(x,\tau)$, considering both single-chain and multi-chain settings. The results are unsatisfactory: for the single chain, $\tau^\mathrm{int}_Q$ remains significantly large even at small $\beta$, while for the multi-chain approach, the topological susceptibility is several times larger than the analytical result.

We also compute the dependence of the acceptance rate on lattice size and different coupling values. Fig.~\ref{fig:acc} shows that the acceptance rate does not decrease significantly as the lattice size and $\beta$ increase.

\section{Analytical Results}
\label{app:exact}

The action of the two-dimensional U(1) gauge field theory can be written as
\begin{equation}
S(U_{\Box}) = -\beta \sum_{\Box}\text{Re}(U_{\Box}) = -\frac{\beta}{2} \sum_{\Box} [U_{\Box} + U_{\Box}^{-1}].
\end{equation}
The partition function is defined as
\begin{align}
    Z = \int dU_{\Box}\, e^{-S(U_{\Box})}
      = \int \prod_{\Box} dU\, e^{\frac{\beta}{2}[U + U^{-1}]}
      = \prod_{\Box} \int \frac{d\phi}{2\pi}\, e^{\beta \cos{\phi}}
      = [I_0(\beta)]^V,
\end{align}
where we used that the Haar measure with the defining representation~\cite{Kanwar:2021wzm}, $U=e^{i\phi}$, is $d\phi/(2\pi)$. The integral is the modified Bessel function of the first kind, $I_{n=0}(\beta)$.

The partition function in fixed topology can also be derived from the known partition function in the $\theta-$vacuum~\cite{Kovacs:1995nn,Bonati:2019ylr,Albandea:2021lvl}. At sufficiently large volume ($V\rightarrow\infty$),
\begin{align}
    Z_Q & = \int \frac{d\phi}{2\pi}\, e^{\beta \sum_{\Box}\cos{\phi}} \delta\left(\sum_{\Box} \frac{\phi}{2\pi} - Q\right)
\nonumber\\
    &    = \int_{-\infty}^{\infty} dk\, e^{-ik Q} \prod_{\Box} I_{k/(2\pi)}(\beta)
     = \int_{-\pi}^{\pi}  d\theta\, e^{-i\theta Q}[I_{\theta/(2\pi)}(\beta)]^V.
\end{align}
In the second line we changed variables, which makes the integrated function decays rapidly beyond the interval $[-\pi, \pi]$. From the partition function, one can derive the probability density over $Q$,
\begin{equation}
    p(Q) \equiv \frac{Z_Q}{Z} = \int_{-\pi}^{\pi}  d\theta\, e^{-i\theta Q}\left[ \frac{I_{\theta/(2\pi)}(\beta)}{I_0(\beta)} \right]^V.
\end{equation}
The susceptibility of the topological charge, $\chi_Q\equiv \langle Q^2 \rangle /V$, is given by
\begin{align}
    \chi_Q = \frac{1}{V} \int dQ\, p(Q) Q^2,
\end{align}
where $Q$ varies over all integers. These expressions are consistent with Refs.~\cite{Kovacs:1995nn,Bonati:2019ylr,Bonati:2019olo,Albandea:2021lvl}.

On $L\times L$ lattices with periodic boundary conditions, the average value of the Wilson loops for U(1) gauge fields can be derived using integration over the set of plaquette variables after gauge fixing~\cite{Kanwar:2021wzm},
\begin{align}
    \langle W_A \rangle = \frac{1}{Z} \sum_k I_k(\beta)^{V - A}\left(\int dU_{\Box}\, U^{k+1} e^{\frac{\beta}{2}[U + U^{-1}]}  \right)
    = \frac{\sum_k I_k(\beta)^{V-A} I_{k+1}(\beta)^A}{\sum_k I_k(\beta)^V},
\end{align}
where $A$ indicates the area of the plaquette variables.

\section{Likelihood Estimation}
\label{app:likelihood_estimation}

Samplers based on stochastic differential equations incorporate a noise term, making it impossible to directly estimate the likelihood of generated samples. In contrast, the probability flow of an ordinary differential equation (ODE) \cite{song2020score} enables such likelihood estimation, as introduced in Ref.~\cite{Wang:2023exq}. The ODE sampler is sufficient for generating configurations at $\beta=1$. By comparing the log-likelihood of samples generated by the trained diffusion model with the physical action, we can assess whether the model has successfully captured the underlying physical distribution.

\begin{figure}[hbpt!]
\begin{center}
    \includegraphics[height=1.8in,width=1.8in]{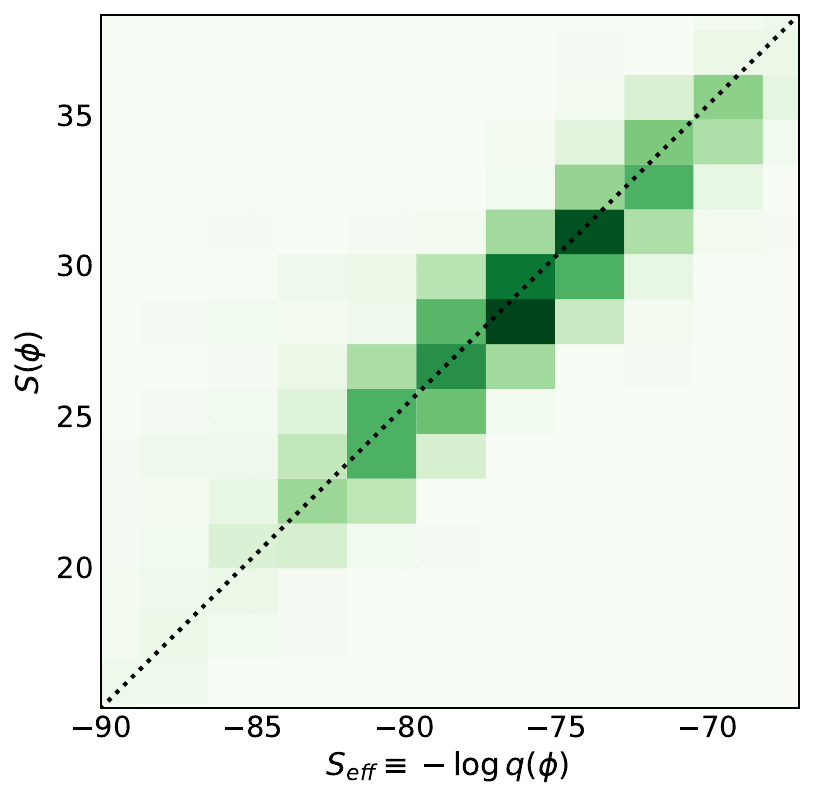}
    \includegraphics[height=1.8in,width=1.8in]{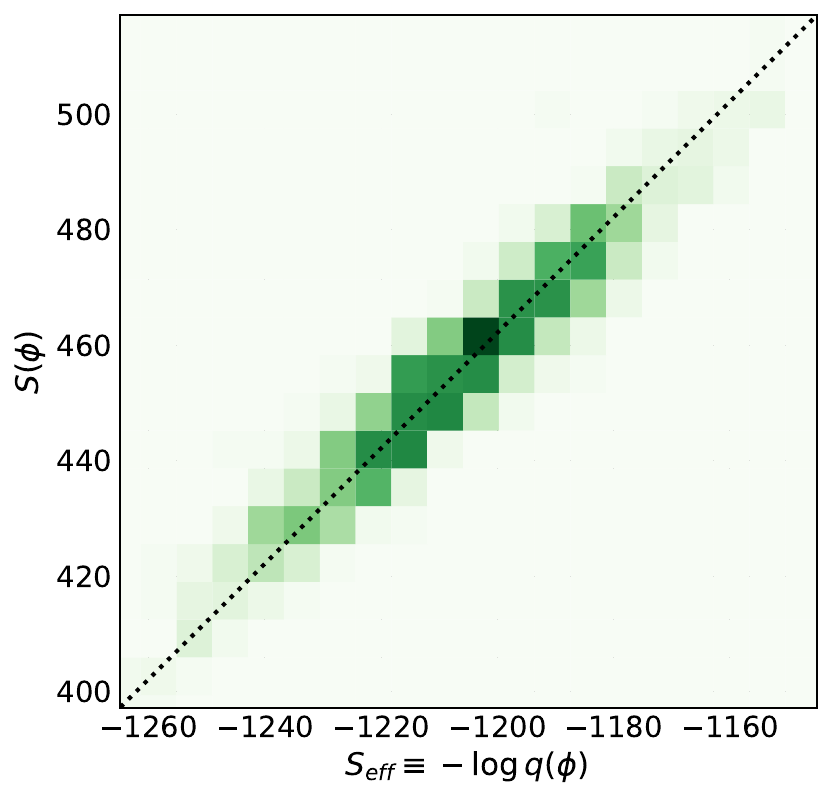}
    \includegraphics[height=1.8in,width=1.8in]{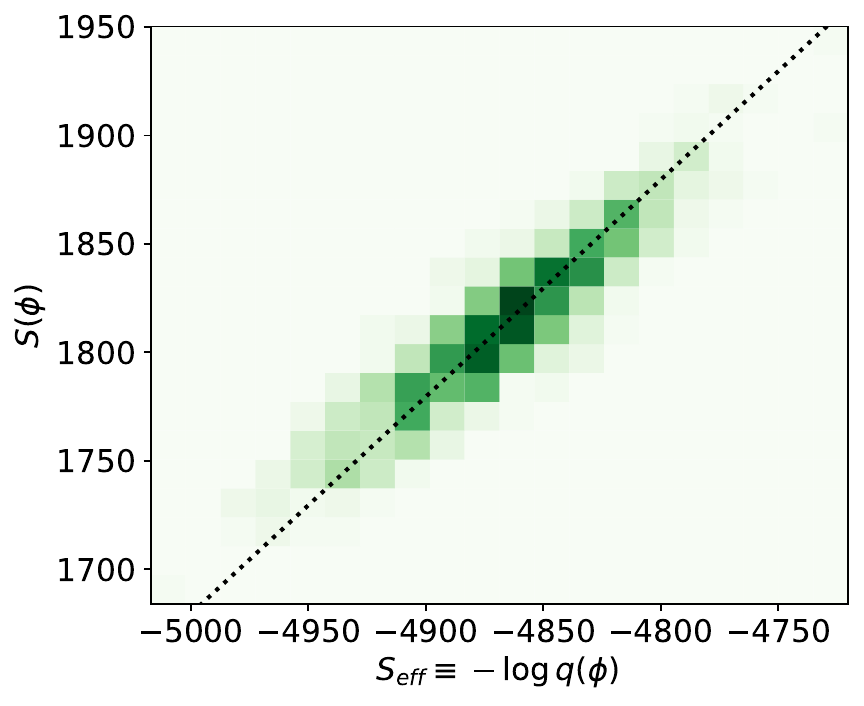}
\end{center}
\caption{
    Likelihood of generated samples at $L=8$ (left), $32$ (middle) and $64$ (right). The vertical axis represents the true action of the configurations, while the horizontal axis corresponds to the log-likelihood generated by the diffusion model, obtained from the probability flow ODE.
}
\label{fig:beta-1}
\end{figure}

\el\noindent
\textbf{ODE sampler and estimator.} To estimate the log-likelihood as if the samples are generated by the probability flow ODE, we use the RK45 ODE solver provided by \\
\texttt{scipy.integrate.solve\_ivp} with \texttt{atol=1e-6} and \texttt{rtol=1e-6}. To generate configurations at $\beta$ and estimate the likelihood, the starting point $\tilde{\mathbf{\phi}}_0\sim \mathcal{N}(0, \sigma_t^2 I)$ and the time interval starts from 1 and ends with 0. The ODE and log-likelihood are computed through
 \begin{equation}
     \frac{d}{dt} \left[ 
     \begin{array}{c}
         \mathbf{\phi}(t) \\
         \log p(\mathbf{\phi}(t), t)
     \end{array}
     \right] 
     = \frac{1}{2} \left[
     \begin{array}{c}
     -  g^2 \cdot s_\theta(\mathbf{\phi}(t), t) \\
     - \sum_{\phi}   g^2 \cdot \frac{\partial s_\theta(\phi(t), t)}{\partial \phi}
     \end{array}
     \right].
\label{eq:ODE}
\end{equation}
In Fig.~\ref{fig:beta-1} we show the correlation of the log-likelihood estimated by the DM, denoted by $S_{\rm eff}=-\log(q(\phi))$, and the real action $S(\phi)$.

\begin{figure}[hbpt!]
\centering
\includegraphics[width = 0.5\textwidth]{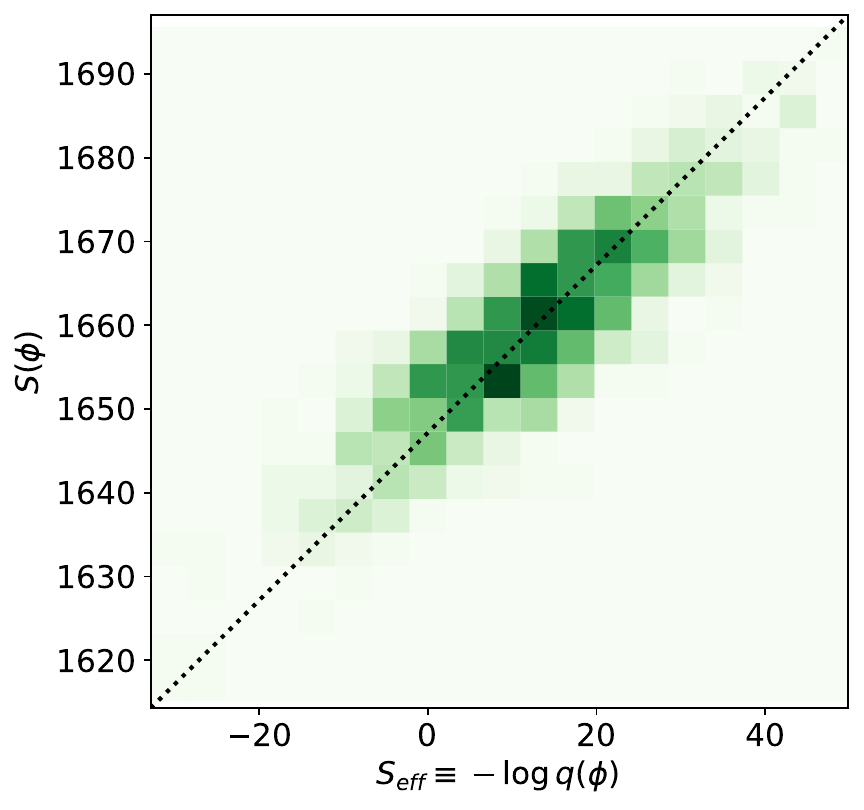}
\caption{
    Likelihood of generated samples at $L=16, \beta = 7$. The vertical axis represents the true action of the configurations, while the horizontal axis corresponds to the log-likelihood generated by the diffusion model, obtained from the probability flow ODE.
}
\label{fig:likelihood-beta-7}
\end{figure}

For $\beta = 7$, since the backward SDE has been modified due to the physics-conditioned design, there is no straightforward method to directly estimate the likelihood as described above. To address this, we trained a new model using 30,720 samples at $\beta = 7$ generated by Algorithm~\ref{alg:annealed-langevin} and applied the same data augmentation method as outlined in Sec.~\ref{sec:numerical_experiments}. Using this newly trained model and Eq.~(\ref{eq:ODE}), we can evaluate the quality of samples generated by Algorithm~\ref{alg:MALAannealed-langevin}. Fig.~\ref{fig:likelihood-beta-7} illustrates the relationship between the log-likelihood and the physical action, treating the samples as if they were generated by the new model using the ODE sampler.

\begin{table}[hbpt!] 
    \centering 
    \caption{Comparison of the topological susceptibility and the $1\times 1$ Wilson loop for $L=16$ and $\beta=1, 7, 11$ using various algorithms.
    }
    \label{tab:L}
    \renewcommand{\arraystretch}{1.3}
    \resizebox{\textwidth}{!}{
    \begin{tabular}{|c|c c c c c c c|}
        \hline
        \multirow{2}{*}{coupling ($\beta$)} & \multicolumn{7}{c|}{Topological Susceptibility} \\
        \cline{2-8}
        & p-HMC-Cold & p-HMC-Hot & HMC & DM & MALA-Hot & MALA-Cold &  Exact  \\
        \hline
        1  & 0.0406(19) &0.0398(18)  & 0.0416(16)       & 0.0422(17)  & 0.0421(20) & 0.0404(18) & 0.0406  \\
        7  & 0.0008(1)  &0.0095(4)   & 0.00013(2)       & 0.0063(3)   & 0.0131(5)  & 0          & 0.0039     \\
        11 & 0          &0.0148(6)   & 0                & 0.0058(3)   & 0.0165(13) & 0          & 0.0024   \\
        \hline
    \end{tabular}
    }
\mbox{}
\vspace*{0.2cm}
\\
    \centering 
    \label{tab:L1}
    \renewcommand{\arraystretch}{1.3}
    \resizebox{\textwidth}{!}{
    \begin{tabular}{|c|c c c c c c c|}
        \hline
        \multirow{2}{*}{coupling ($\beta$)} & \multicolumn{7}{c|}{$1 \times 1$ Wilson Loop} \\
        \cline{2-8}
        & p-HMC-Cold & p-HMC-Hot & HMC & DM & MALA-Hot & MALA-Cold &  Exact  \\
        \hline
        1  & 0.445(38)  &0.4478(38) & 0.447(37)  & 0.446(37)  & 0.444(36) & 0.444(37) & 0.446  \\
        7  & 0.927(7)   &0.925(7)   & 0.926(7)   & 0.926(7)   & 0.924(6)  & 0.925(7)  & 0.926     \\
        11 & 0.954(4)   &0.953(5)   & 0.954(3)   & 0.953(4)   & 0.950(5)  & 0.952(4)  & 0.953   \\
        \hline
    \end{tabular}
    }
\end{table}

\section{Different Sampler Results}
\label{app:para}


Here we provide a more apples-to-apples comparison between the various algorithms. In Table \ref{tab:L} we show results for the topological susceptibility and the $1\times 1$ Wilson loop. We compare:

\begin{itemize}
\item parallel HMC (p-HMC) with cold/hot starts (cold means the initial state is all zeroes; hot means the initial state is drawn from a standard Gaussian distribution); 
\item
single-chain hybrid Monte Carlo (HMC);
\item
diffusion model (DM), including accept/reject and annealing. The reported DM results are the ‘ensemble-averaged’ outcomes: the mean over the models from the last eight training steps, each evaluated on 128 configurations. One could also consider Bayesian model averaging as an better alternative.; 
\item
Langevin dynamics including the accept/reject step (MALA), using cold/hot starts, all other setting is consistent with DM sampler in Appendix \ref{app:langevin} except that the learned drift is replaced by the gradient of the target action. 
\end{itemize}
For HMC and MALA a burn-in stage is always included before Metropolis is employed. 
Since at large $\beta$ every chain in HMC and MALA is frozen, the final result will depend more and more on the initial state. Hence for a hot start, the topological susceptibility at $\beta=11$ is larger than at $\beta=7$. 
The diffusion model is not sensitive to the initial state and topological sectors can be explored sufficiently well.

\bibliographystyle{JHEP}
\bibliography{ref}

\end{document}